\documentclass[manuscript]{aastex}

\usepackage{hyperref}
\usepackage{url}
\usepackage{lscape}

\begin{document}

\title{Anisotropic Star on Pseudo-Spheroidal Spacetime}

\author{B. S. Ratanpal\altaffilmark{1}}
\affil{Department of Applied Mathematics, Faculty of Technology \& Engineering, The M. S. University of Baroda, Vadodara - 390 001, India}
\email{bharatratanpal@gmail.com}

\author{V. O. Thomas\altaffilmark{2}}
\affil{Department of Mathematics, Faculty of Science, The M. S. University of Baroda,\\ Vadodara - 390 002, India}
\email{votmsu@gmail.com}

\and

\author{D. M. Pandya\altaffilmark{3}}
\affil{Department of Mathematics \& Computer Science, Pandit Deendayal Petroleum University, Raisan, Gandhinagar - 382 007, India}
\email{dishantpandya777@gmail.com}

\begin{abstract}
A new class of exact solutions of Einstein's field equations representing anisotropic distribution of matter on
pseudo-spheroidal spacetime is obtained. The parameters appearing in the model are restricted through physical
requirements of the model. It is found that the models given in the present work is compatible with observational data of 
a wide variety of compact objects like 4U 1820-30, PSR J1903+327, 4U 1608-52, Vela X-1, PSR J1614-2230, SMC X-4, Cen X-3. A 
particular model of pulsar PSR J1614-2230 is studied in detail and found that it satisfies all physical requirements needed
for physically acceptable model.
\end{abstract}


\section{Introduction}
The study of interior solutions of Einstein's field equations play a significant role in predicting the nature of the star in the last stages of evolution. There is an emerging interest among researchers to develop mathematical models of superdense stars that are compatible with observational data. A number of articles appeared recently in literature that are in good agreement with observational data \cite{Pandya14,Gangopadhyay13,Hansraj06,Tik09,Ayan13}. The spacetime metric representing these models may not have a known 3-space geometry. Many researchers used spacetimes having known 3-space geometry to develop superdense star models. The spheroidal spacetime used by \cite{Vaidya82} \cite{Tikekar90} and the paraboloidal spacetime used by \cite{Finch89}, \cite{Jotania07}, \cite{Sharma13},\cite{Pandya14} are examples of spacetimes with definite 3-space geometry. The spacetime metric used by \cite{Tikekar98}, \cite{Thomas05}, \cite{Thomas07}, \cite{Jotania05},\cite{Chattopadhyay10} spheroidal geometry for its physical three space. It is found that all the above spacetimes are useful to describe compact objects like 
neutron stars and quark stars. \\

The matter content of astrophysical objects in ultra high densities may not be in the form of perfect fluid. Theoretical study of \cite{Ruderman72} and \cite{Canuto74} about more realistic stars suggest that in a very high density regime, matter may not be isotropic. Anisotropy arises due to the existence of solid stellar core or by presence of type-3A superfluid \cite{Kippenhahn90}, \cite{Sokolov80}, phase transitions \citep{Sawyer72} or pion condensation in a star. Study of anisotropic distributions of matter got wide attention after the pioneering work of \cite{Bowers74}. A number of articles appeared in literature related to compact anisotropic stars. The anisotropic model developed by \cite{Maharaj89} have uniform density for its matter content. \cite{Gokhroo94} gave a 
more realistic model incorporating non-uniform density. \cite{Tikekar98,Tikekar05}, \cite{Thomas05} developed superdense anisotropic 
distributions on pseudo-spheroidal spacetime. The spacetime used to study the non-adiabatic gravitational collapse of anisotropic 
distribution has a pseudo spheroidal geometry. \cite{Dev02,Dev03,Dev04} have study the impact of anisotropy on the stability of stars.
Anisotropic distributions with linear equation of state have been studied by \cite{Sharma07}, \cite{Thirukkanesh08}. \cite{Komathiraj07} studied charged distributions using linear equation of state. \cite{Sunzu14} used linear equation of state for describing charged quark star models. \cite{Feroze11} and \cite{Maharaj12} developed 
models of anisotropic distributions using quadratic equation of state. In the MIT bag model of quark stars, \cite{Paul11}
have shown that introduction of anisotropy can affect bag constant. \cite{Thirukkanesh12}, \cite{Maharaj13b} used polytropic
equation of state for developing anisotropic models. The anisotropic charged star models given by \cite{Malaver113,Malaver213,Malaver14} and \cite{Thirukkanesh14}
have Van der Waals equation of state. \cite{Pandya14} have given anisotropic compact stars compatible with observational
data by generalizing \cite{Finch89} model. The anisotropic model given by \cite{Sharma13} is a particular case of this model.
It is found that the models given by \cite{Pandya14} agree with the recent observational data given by \cite{Gangopadhyay13}.

In this paper we have studied anisotropic stellar models on pseudo-spheroidal spacetime. The geometric parameter $R$ plays the role of 
radius of the star. The geometric parameter $K$ appearing in the model is bounded between two limits to comply with the physical
requirements. It is found that the anisotropic model developed can accommodate wide range of pulsars like 4U 1820-30, 
PSR J1903+327, 4U 1608-52, Vela X-1, PSR J1614-2230, SMC X-4, Cen X-3.

In section \ref{sec:2}, we have given the spacetime metric and obtained solution of Einstein's field equations. The bounds of model parameter
$K$ is obtained in section \ref{sec:3} by imposing the physical requirements a physically viable model is expected to satisfy. In
section \ref{sec:4}, we have shown that the anisotropic compact star models are in agreement with the observational data of 
\cite{Gangopadhyay13} and discuss the main results obtained at the end. 

\section{Space-time Metric}
\label{sec:2}
We shall take the pseudo-spheroidal spacetime metric for describing the anisotropic matter distribution in the form
\begin{equation}\label{IMetric1}
	ds^{2}=e^{\nu(r)}dt^{2}-\left(\frac{1+K\frac{r^{2}}{R^{2}}}{1+\frac{r^2}{R^{2}}} \right)dr^{2}-r^{2}\left(d\theta^{2}+\sin^{2}\theta d\phi^{2} \right),
\end{equation}
where $K,\;R$ are geometric parameters and $K>1$.
The energy-momentum tensor for the anisotropic matter distribution \citep{Maharaj89} is taken as
\begin{equation}\label{EMTensor}
	T_{ij}=\left(\rho+p\right)u_{i}u_{j}-pg_{ij}+\pi_{ij},
\end{equation}
where $\rho$, the proper density, $p$ denotes fluid pressure and $u_{i}$ denotes the unit four-velocity of the fluid. The anisotropic 
stress tensor $\pi_{ij}$ is given by
\begin{equation}\label{Pi}
	\pi_{ij}=\sqrt{3}S\left[c_{i}c_{j}-\frac{1}{3}\left(u_{i}u_{j}-g_{ij} \right) \right].
\end{equation}
Here $S=S(r)$ is the magnitude of the anisotropic stress and $c^{i}=\left(0,-e^{-\lambda/2},0,0 \right)$ denotes a radial vector.

We have now the following expressions for radial and transverse pressures
\begin{equation}\label{pr1}
	p_{r}=-T_{1}^{1}=p+\frac{2S}{\sqrt{3}},
\end{equation}
and
\begin{equation}\label{pp1}
	p_{\perp}=-T_{2}^{2}=p-\frac{S}{\sqrt{3}}.
\end{equation}
The difference in radial and transverse pressures is taken as the measure of anisotropy and is given by
\begin{equation}\label{S}
	S=\frac{p_{r}-p_{\perp}}{\sqrt{3}}.
\end{equation}
The Einstein's field equations
\begin{equation}\label{EFE}
	R_{ij}-\frac{1}{2}Rg_{ij}=8\pi T_{ij},
\end{equation}
are equivalent to following set of three equations
\begin{equation}\label{rho2}
	8\pi\rho=\frac{1-e^{-\lambda}}{r^{2}}+\frac{e^{-\lambda}\lambda'}{r},
\end{equation}
\begin{equation}\label{pr2}
	8\pi p_{r}=\frac{e^{-\lambda}-1}{r^{2}}-\frac{e^{-\lambda}\nu'}{r},
\end{equation}
\begin{equation}\label{pp2}
	8\pi p_{\perp}=e^{-\lambda}\left(\frac{\nu''}{2}+\frac{\nu'^{2}}{4}-\frac{\nu'\lambda'}{4}+\frac{\nu'-\lambda'}{2r} \right).
\end{equation}
Equation (\ref{rho2})-(\ref{pp2}) can be further modified to the following form,
\begin{equation}\label{FE1}
	e^{-\lambda}=1-\frac{2m}{r},
\end{equation}
\begin{equation}\label{FE2}
	\left(1-\frac{2m}{r} \right)\nu'=8\pi p_{r}r+\frac{2m}{r^{2}},
\end{equation}
\begin{equation}\label{FE3}
	-\frac{4}{r}\left(8\pi\sqrt{3}S \right)=\left(8\pi\rho+8\pi p_{r} \right)\nu'+2\left(8\pi p_{r}' \right),
\end{equation}
where,
\begin{equation}\label{m}
	m(r)=4\pi\int_{0}^{r}u^{2}\rho(u)du.
\end{equation}
Using
\begin{equation}\label{g11}
	e^{\lambda}=\frac{1+K\frac{r^{2}}{R^{2}}}{1+\frac{r^{2}}{R^{2}}},
\end{equation}
in (\ref{rho2}), we obtain the expression for $\rho$ in the form
\begin{equation}\label{rho3}
	8\pi\rho=\frac{K-1}{R^{2}}\frac{3+K\frac{r^{2}}{R^{2}}}{\left(1+K\frac{r^{2}}{R^{2}} \right)^{2}}.
\end{equation}

The metric potential $\nu$ can be obtained from equation (\ref{FE2}) once we know the expression for $p_{r}$. We shall take
the expression for radial pressure as
\begin{equation}\label{pr3}
	8\pi p_{r}=\frac{K-1}{R^{2}}\frac{1-\frac{r^{2}}{R^{2}}}{\left(1+K\frac{r^{2}}{R^{2}} \right)^{2}}.
\end{equation}
It can be noticed from equation (\ref{pr3}) that the central pressure is $p_{r}(0)=\frac{K-1}{R^2}$, which is directly related to the 
geometric parameters $K$ and $R$. This choice of $p_{r}$ facilitate the integration of equation (\ref{FE2}) and obtain $e^{\nu}$ in the form
\begin{equation}\label{enu}
	e^{\nu}=CR^{\frac{K^{2}-2K+1}{K}}\left(1+K\frac{r^{2}}{R^{2}} \right)^{\frac{K+1}{2K}}\left(1+\frac{r^{2}}{R^{2}} \right)^{\frac{K-3}{2}},
\end{equation}
where $C$ is a constant of integration.

Differentiating equation (\ref{pr3}) with respect to r, we get
\begin{equation}\label{dprdr}
	8\pi\frac{dp_{r}}{dr}=-\frac{2r(K-1)}{R^{4}}\frac{1+K\left(2-\frac{r^{2}}{R^{2}} \right)}{\left(1+K\frac{r^{2}}{R^{2}} \right)^{3}}.
\end{equation}
It can be noticed from equation (\ref{dprdr}) that
\begin{equation}\label{dprdr0}
	8\pi\frac{dp_{r}}{dr}\left(r=0 \right)=0,
\end{equation}
\begin{equation}\label{dprdrR}
	8\pi\frac{dp_{r}}{dr}\left(r=R \right)=-\frac{2(K-1)}{R^2(1+K)^{2}}< 0.
\end{equation}
Further $8\pi\frac{dp_{r}}{dr}<0$ for all values of $r$ in the range $0\leq r\leq R$. Hence the radial pressure is decreasing radially
outward and becomes zero at $r=R$, which is taken as the radius of the anisotropic fluid distribution.

From equation (\ref{rho3}), we get
\begin{equation}\label{drhodr}
	8\pi \frac{d\rho}{dr}=-\frac{2rK(K-1)}{R^{4}}\frac{5+K\frac{r^{2}}{R^{2}}}{\left(1+K\frac{r^{2}}{R^{2}} \right)^{3}}<0,
\end{equation}
indicating that the density is a decreasing function of $r$.

The spacetime metric (\ref{IMetric1})now takes the explicit form
\begin{eqnarray}\label{IMetric2}
	ds^{2} & = & CR^{\frac{K^2-2K+1}{k}}\left(1+K\frac{r^{2}}{R^{2}} \right)^{\frac{K+1}{2K}}\left(1+\frac{r^{2}}{R^{2}} \right)^{\frac{k-3}{2}}dt^{2}-\left(\frac{1+K\frac{r^{2}}{R^{2}}}{1+\frac{r^{2}}{R^{2}}} \right)dr^{2}\\\nonumber
	       &   & -r^{2}\left(d\theta^{2}+\sin^{2}\theta d\phi^{2} \right).
\end{eqnarray}

For a physically acceptable relativistic distribution of matter, the interior spacetime metric (\ref{IMetric2}) should continuously match
with Schwarzschild exterior metric
\begin{equation}\label{EMetric1}
	ds^{2}=\left(1-\frac{2M}{r} \right)dt^{2}-\left(1-\frac{2M}{r} \right)^{-1}dr^{2}-r^{2}\left(d\theta^{2}+\sin^{2}\theta d\phi^{2} \right),
\end{equation}
across the boundary $r=R$. This gives the constants $M$ and $C$ as
\begin{equation}\label{M}
	M=\frac{R}{2}\frac{\left(K-1\right)}{\left(K+1\right)},
\end{equation}
and
\begin{equation}\label{C}
	C=R^{-\left(\frac{K^2-2K+1}{K} \right)}\left(1+K \right)^{-\left(\frac{3K+1}{2K} \right)}2^{\left(\frac{5-K}{2} \right)}.
\end{equation}

The expression for anisotropy $S$ can be obtained using (\ref{rho3}), (\ref{pr3}), (\ref{enu}) and (\ref{dprdr}) in (\ref{FE3}). We have
\begin{equation}\label{S1}
	8\pi\sqrt{3}S=\frac{(K-1)r^{2}\left[12+\left(-6K^2+16K-2 \right)\frac{r^{2}}{R^{2}}+\left(-K^{3}+3K^2-7K+1\right)\frac{r^{2}}{R^{2}} \right]}{4R^{4}\left(1+K\frac{r^{2}}{R^{2}} \right)^{3}\left(1+\frac{r^{2}}{R^{2}} \right)},
\end{equation}
which takes the value zero at $r=0$. The expression for $8\pi p_{\perp}=8\pi p_{r}-8\pi\sqrt{3}S$, now takes the form
\begin{equation}\label{pp3}
	8\pi p_{\perp}=\frac{\left(K-1 \right)\left[4+\left(4K-12 \right)\frac{r^{2}}{R^{2}}+\left(6K^2-16K-2 \right)\frac{r^{4}}{R^{4}}+\left(K^3-3K^2+3K-1 \right)\frac{r^{6}}{R^{6}} \right]}{4R^{2}\left(1+K\frac{r^{2}}{R^{2}} \right)^{3}\left(1+\frac{r^{2}}{R^{2}} \right)},
\end{equation}
\section{Physical Analysis}
\label{sec:3}
The anisotropic matter distribution described on the background of pseudo-spheroidal spacetime contains two geometric parameters
$R$ and $K$. Since $p_{r}(r=R)=0$, the parameter $R$ represents the radius of the distribution. The bounds on the other parameter
$K$ is to be determined using the physical plausibility conditions stipulated below:
\begin{itemize}
\item [(i)] $\rho(r),~p_{r}(r),~p_\perp(r) \geq 0 $ for $ 0 \leq r \leq R$;
\item [(ii)] $\rho-p_{r}-2p_\perp \geq 0$ for $ 0 \leq r \leq R$;
\item [(iii)] $\frac{d\rho}{dr},~\frac{dp_{r}}{dr},~\frac{dp_{t}}{dr} < 0$ for $0 \leq r \leq R$;
\item [(iv)] $0 \leq \frac{dp_{r}}{d\rho} \leq 1$; $0 \leq \frac{dp_\perp}{d\rho} \leq 1$, for $0 \leq r \leq R$;
\item [(v)] The adiabatic index $\Gamma(r)>\frac{4}{3}$, for $0 \leq r \leq R$.
\end{itemize}
The conditions $\rho(r)\geq 0$, $p_{r}\geq 0$, $\frac{d\rho(r)}{dr}<0$ and $\frac{dp_{r}}{dr}<0$ are evidently satisfied in the light of 
equations (\ref{rho3}), (\ref{pr3}), (\ref{drhodr}) and (\ref{dprdr}), respectively.

The condition $p_{\perp}>0$ imposes a restriction on the value of $K$, namely,
\begin{equation}\label{K1}
	K\geq 2.4641.
\end{equation}

In order to examine the strong energy condition, we evaluate the expression $\rho-p_{r}-2p_{\perp}$ at $r=0$ and at $r=R$. It is easy to see that
\begin{equation}\label{SEC1}
	\left(\rho-p_{r}-2p_{\perp} \right)(r=0)=0,
\end{equation}
and $\left(\rho-p_{r}-2p_{\perp} \right)(r=R)\geq0$, imposes an upper bound for $K$, namely,
\begin{equation}\label{K2}
	K\leq 4.1231.
\end{equation}
It has been suggested by \cite{Canuto74} that he velocity of sound should be monotonically decreasing for matter distribution with ultra-high
densities. This demands that $\frac{d}{dr}\left(\frac{dp_{r}}{d\rho} \right)<0$.

The expressions for $\frac{dp_{r}}{d\rho}$ and $\frac{dp_{\perp}}{d\rho}$ are given by
\begin{equation}\label{dprdrho}
	\frac{dp_{r}}{d\rho}=\frac{1+2K-K\frac{r^{2}}{R^{2}}}{K\left(5+K\frac{r^{2}}{R^{2}} \right)},
\end{equation}

It can be noticed from (\ref{dprdrho}) that $ \frac{dp_r}{d\rho} \leq 1 $ throughout the distribution.

\begin{equation}\label{dppdrho}
	\frac{dp_{\perp}}{d\rho}=\frac{\left(1+K\frac{r^{2}}{R^{2}} \right)^{3}\left[X_{1}+X_{2}\frac{r^{2}}{R^{2}}+X_{3}\frac{r^{4}}{R^{4}}+X_{4}\frac{r^{6}}{R^{6}}+X_{5}\frac{r^{8}}{R^{8}} \right]}{2K(K-1)\left(5+K\frac{r^{2}}{R^{2}} \right)\left[2+Y_{1}\frac{r^{2}}{R^{2}}+Y_{2}\frac{r^{4}}{R^{4}}+Y_{3}\frac{r^{6}}{R^{6}}+Y_{4}\frac{r^{8}}{R^{8}}+Y_{5}\frac{r^{10}}{R^{10}}+2K^{4}\frac{r^{12}}{R^{12}} \right]},
\end{equation}
where, $X_{1}=8K^{2}+8K-16$, $X_{2}=-4K^3+28K^2-20K-4$, $X_{3}=3K^4-4K^3-30K^2+36K-5$, $X_{4}=10K^4-36K^3+16K^2+12K-2$, $X_{5}=K^{5}-4K^{4}+6K^{3}-4K^{2}+4$
and $Y_{1}=8K+4$, $Y_{2}=12K^2+16K+2$, $Y_{3}=8K^{3}+24K^{2}+8K$, $Y_{4}=2K^{4}+16K^{3}+12K^{2}$, $Y_{5}=4K^{4}+8K^{3}$.
 
The condition $\frac{dp_{\perp}}{d\rho}\leq 1$ at $r=0$ and $r=R$ gives the following bounds on $K$, viz.,
\begin{equation}\label{K5}
	K>1.3333,
\end{equation}
and
\begin{equation}\label{K6}
	1\leq K\leq 14.7882.
\end{equation}
The expression for adiabatic index $ \Gamma $ is given by

\begin{equation}
 \Gamma = \frac{\left(4 - \frac{r^2}{R^2} + K \frac{r^2}{R^2}\right)\left(1 + 2K - K \frac{r^2}{R^2}\right)}{K\left(1 - \frac{r^2}{R^2}\right)\left(5 + K \frac{r^2}{R^2}\right)}
 \label{gamma}
\end{equation}

The necessary condition for the model to represent a relativistic star is that $\Gamma>\frac{4}{3}$ throughout the star. $\Gamma>\frac{4}{3}$
at $r=0$ imposes a condition on $K$, viz., 
\begin{equation}\label{K6}
	K>-3.
\end{equation}
Considering all the relevant inequalities, we find that the admissible bound for $K$ is given by
\begin{equation}\label{K8}
	2.4641\leq K\leq 4.1231.
\end{equation}
\section{Application to Compact Stars and Discussion}
\label{sec:4}
In order to examine the suitability of our model to fit into the observational data, we have considered the masses and radii of some well
known pulsars given by \cite{Gangopadhyay13}. We have considered PSR J1614-2230 whose estimated mass and radius are $1.97 M_{\odot}$ and $9.69\; km$.
If we set these values in equation (\ref{M}) we get $K=3.997$ which is well inside the valid range of $K$. We have further verified that our model
is in good agreement with the estimated mass and radii of a number of compact stars like Vela X-1, 4U1608-52, PSRJ1903+327, 4U1820-30
SMC X-4 and Cen X-3. The value of $K$, mass, radius and other relevant quantities like $\rho_{c}$, $\rho_{R}$, $u=\frac{M}{R}$ and 
$\frac{dp_{r}}{d\rho}(r=0)$ are shown in Table-1.

\begin{table}[h]
\caption{Estimated physical values based on the observational data}
\label{tab:1}
\begin{tabular}{llllllll}
\hline\noalign{\smallskip}
\textbf{STAR} & $ K $ & \textbf{$ M $} & \textbf{$ R $} & \textbf{$ \rho_c $} & \textbf{$ \rho_R $} & \textbf{$ u (=\frac{M}{R}) $} & $\left(\frac{dp_r}{d \rho}\right)_{r=0} $ \\
& & $ (M_\odot) $ & (Km) & (MeV fm{$^{-3}$}) & (MeV fm{$^{-3}$}) & \\
\noalign{\smallskip}\hline\noalign{\smallskip}
\textbf{4U 1820-30} 	  & 3.100 & 1.58  & 9.1   & 2290.97 & 277.12 & 0.256 & 0.465 \\
\textbf{PSR J1903+327} 	  & 3.176 & 1.667 & 9.438 & 2206.89 & 260.52 & 0.261 & 0.463 \\
\textbf{4U 1608-52} 	  & 3.458 & 1.74  & 9.31  & 2561.92 & 277.50 & 0.276 & 0.458 \\
\textbf{Vela X-1} 	  & 3.407 & 1.77  & 9.56  & 2379.27 & 261.63 & 0.273 & 0.459 \\
\textbf{PSR J1614-2230}   & \textbf{3.997} & \textbf{1.97}  & \textbf{9.69}  & \textbf{2883.52} & \textbf{269.33} & \textbf{0.300} & \textbf{0.450} \\
\textbf{SMC X-4}          & 2.514 & 1.29  & 8.831 & 1753.84 & 261.06 & 0.215 & 0.480 \\
\textbf{Cen X-3}          & 2.838 & 1.49  & 9.178 & 1971.21 & 260.42 & 0.239 & 0.470 \\
\noalign{\smallskip}\hline
\end{tabular} 
\end{table}

\begin{figure}
\includegraphics[scale = 1.25]{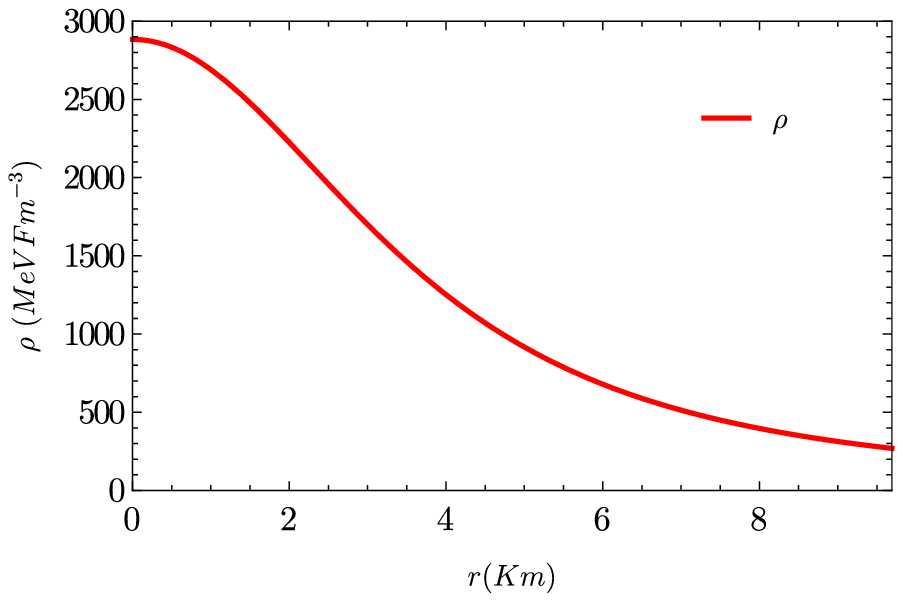}
\caption{Variation of density against radial variable $r$. 
\label{fig1}}
\end{figure}

\begin{figure}
\includegraphics[scale = 1.25]{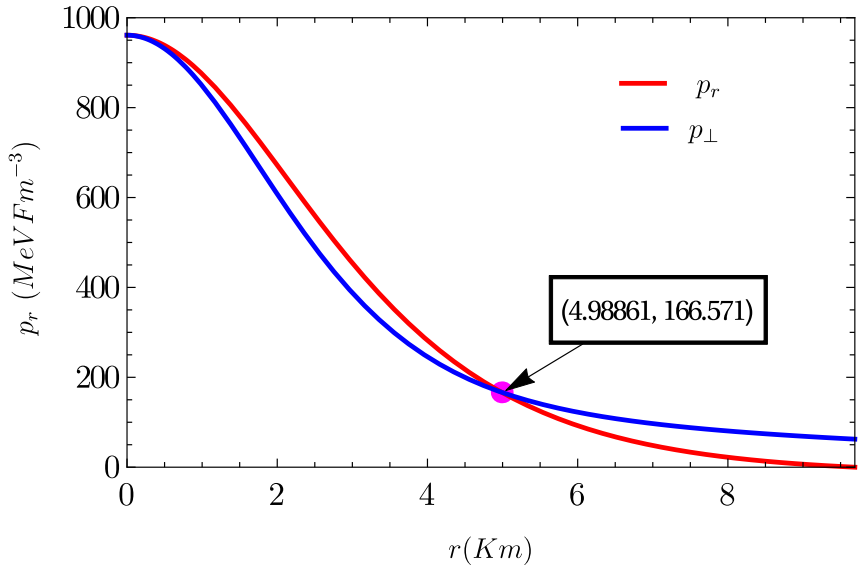}
\caption{Variation of pressures against radial variable $r$. 
\label{fig2}}
\end{figure}

\begin{figure}
\includegraphics[scale = 1.25]{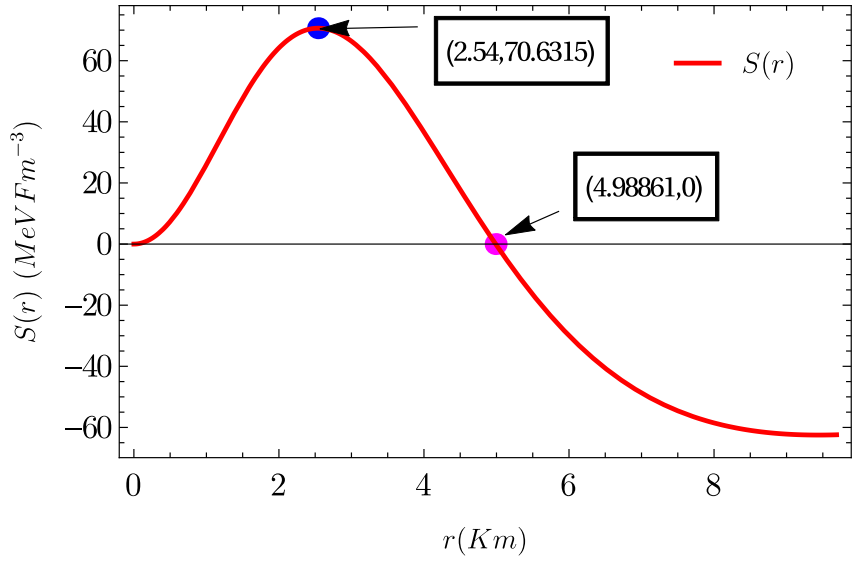}
\caption{Variation of anisotropy against radial variable $r$. 
\label{fig3}}
\end{figure}

\begin{figure}
\includegraphics[scale = 1.25]{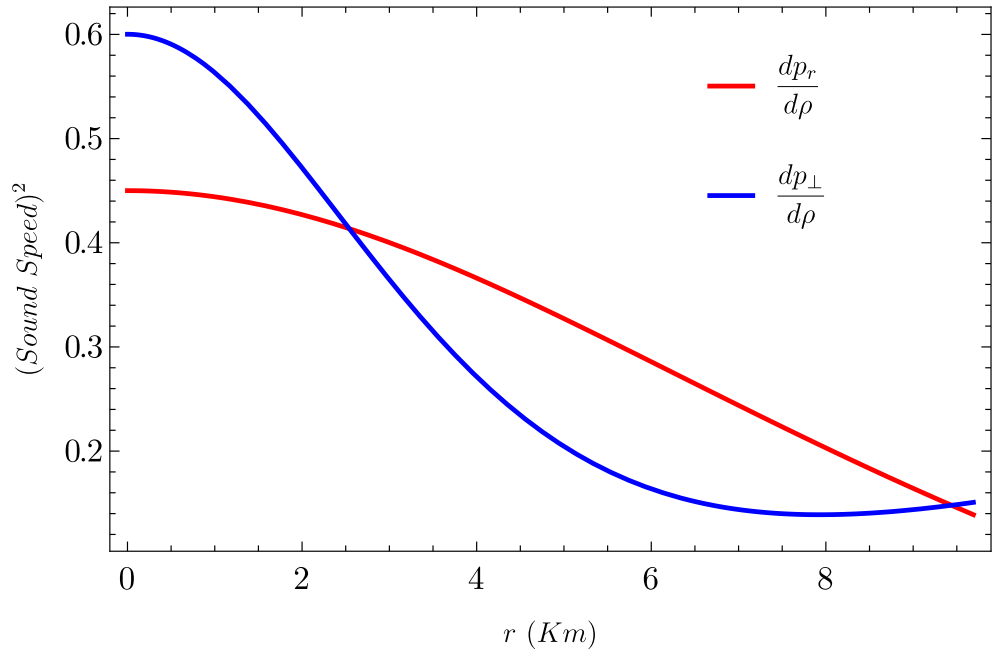}
\caption{Variation of $ \frac{1}{c^2}\frac{dp_r}{d\rho}, ~ \frac{1}{c^2}\frac{dp_\perp}{d\rho}$ against radial variable $r$. 
\label{fig4}}
\end{figure}

\begin{figure}
\includegraphics[scale = 1.25]{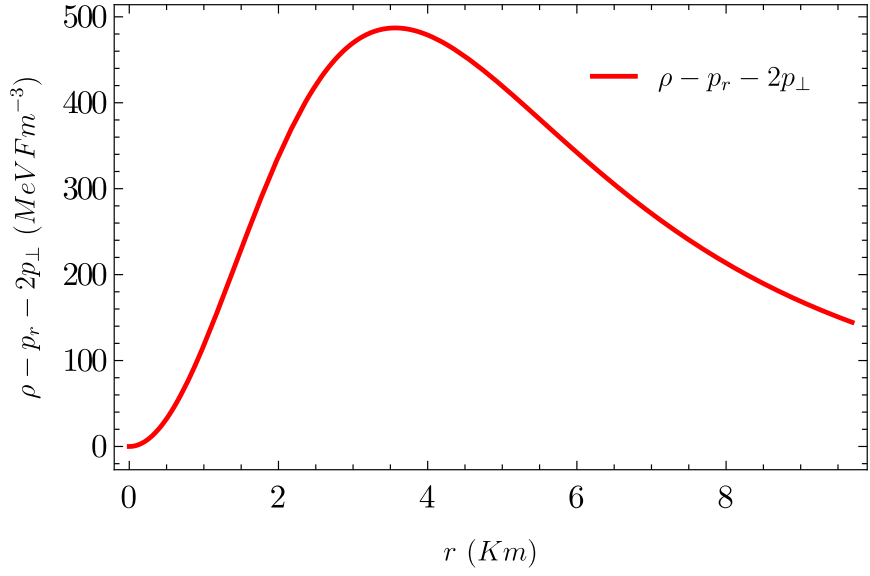}
\caption{Strong energy condition Vs radial variable $r$. 
\label{fig5}}
\end{figure}

In order to have detailed analysis of various physical conditions throughout the star we have considered a particular star 
PSR J1614-2230 having mass $M=1.97M_{\odot}$ and radius $R=9.69\;km$ along with the geometric parameter $K=3.997$. The
variation of density and pressure from centre to the boundary of the star is shown graphically in Figure~\ref{fig1} and Figure~\ref{fig2}, respectively.
It can be seen that density and pressures are monotonically decreasing functions of the radial variable $r$. In Figure~\ref{fig3}, we have 
shown the variation of anisotropy $S$ throughout the star. The anisotropy increases initially and after reaching a maximum at
$r=2.54$, then it starts decreasing till the boundary of the star. The variation of square of sound speed and strong energy condition are 
displayed in Figure~\ref{fig4} and Figure~\ref{fig5}, respectively. It can be noticed that the square of sound speed is less than 1 and the strong energy
condition is satisfied throughout the star.

\begin{figure}
\includegraphics[scale = 1.25]{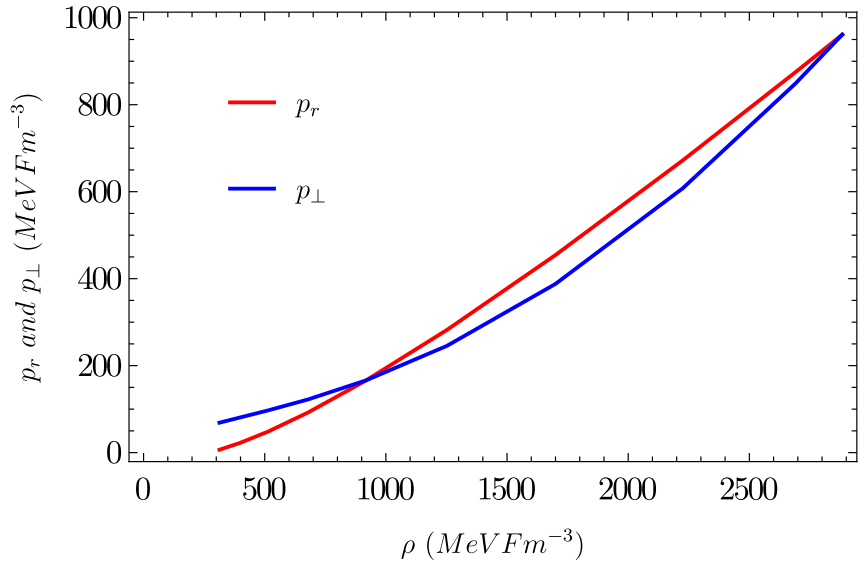}
\caption{Variation of $p_r$ and $p_\perp$ against $\rho$. 
\label{fig6}}
\end{figure}

\begin{figure}
\includegraphics[scale = 1.25]{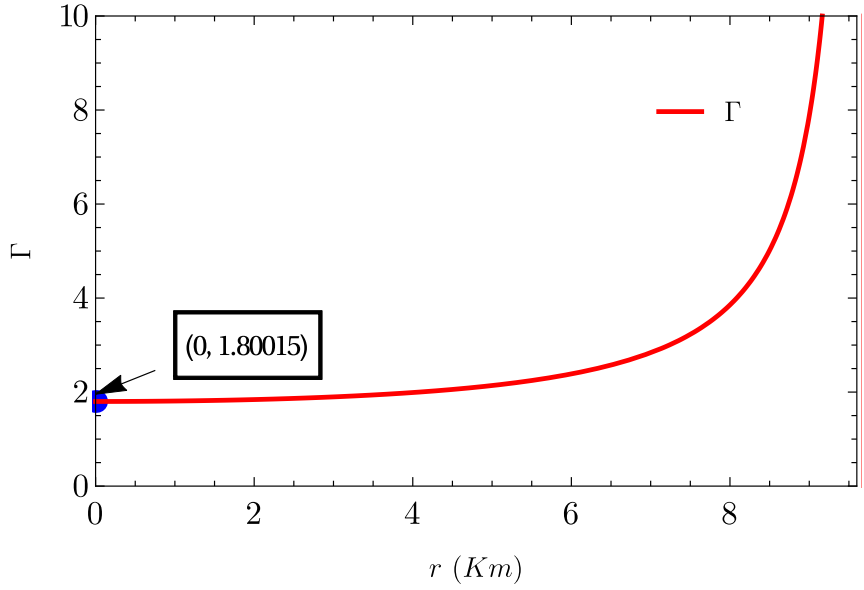}
\caption{Variation of $\Gamma$ against radial variable $r$. 
\label{fig7}}
\end{figure}

\begin{figure}
\includegraphics[scale = 1.25]{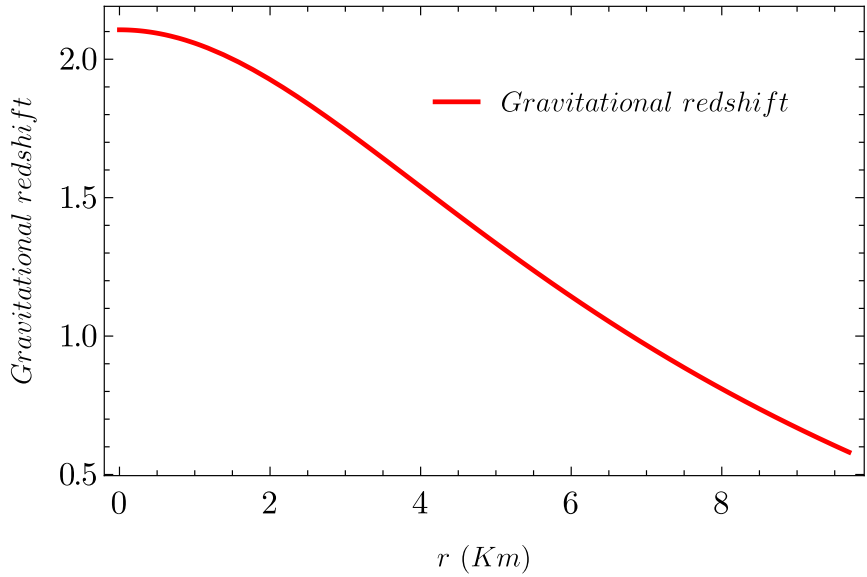}
\caption{Variation of gravitational redshift against radial variable $r$. 
\label{fig8}}
\end{figure}

In Figure~\ref{fig6}, we have shown the equation of state for matter distribution in graphical form. For a stable relativistic star, the 
adiabatic index $\Gamma$ should be greater than $\frac{4}{3}$ throughout the configuration. We have plotted the graph of $\Gamma$
against $r$ in Figure~\ref{fig7}. The graph clearly indicates that $\Gamma>\frac{4}{3}$ throughout the star. For a physically acceptable model,
the gravitational redshift, $z=\sqrt{e^{-\nu(r)}}-1$, should be a decreasing function of $r$. Further the central redshift
$z_{c}$ and boundary redshift $ z_R $ should be positive and finite. From Figure~\ref{fig8}, it can be seen that these conditions are satisfied throughout
the star.

We have studied spherically symmetric anisotropic distributions of matter on pseudo-spheroidal spacetime. The model we have developed
is in good agreement with the observational data of pulsars recently studied by \cite{Gangopadhyay13}. The model parameters
are carefully selected so that the models satisfy all the physical requirements throughout the distribution. We have studied a particular
model of PSR J1614-2230 and have shown that various physical requirements stipulated earlier are satisfied throughout the star.
\section*{Acknowlegement}
The authors would like to thank IUCAA, Pune for the facilities and hospitality provided to them for carrying out this work.

\pagebreak

\end{document}